# Mirror thermal noise in laser interferometer gravitational wave detectors operating at room and cryogenic temperature

Janyce Franc, Nazario Morgado, Raffaele Flaminio, Ronny Nawrodt, Iain Martin, Liam Cunningham, Alan Cumming, Sheila Rowan and James Hough

**Abstract**

Mirror thermal noise is and will remain one of the main limitations to the sensitivity of gravitational wave detectors based on laser interferometers. We report about projected mirror thermal noise due to losses in the mirror coatings and substrates. The evaluation includes all kind of thermal noises presently known. Several of the envisaged substrate and coating materials are considered. The results for mirrors operated at room temperature and at cryogenic temperature are reported.

## 1. Thermal noise

Thermal noise is an important limit to the sensitivity of gravitational waves detectors (GWD) in the frequency range of 50-500 Hz. The Einstein Telescope project aims at reducing the thermal noise affecting GWD to a few in $10^{-25}$ m/√Hz instead of a few $10^{-24}$ m/√Hz planned for the second generation of GWD. Decreasing thermal noise has become an important theoretical and experimental challenge. The thermal noise affecting gravitational wave interferometers has two different origins. The first one is due to dissipation in the wires used to suspend the test masses; this is the so called suspension thermal noise. The second one is due to dissipation processes inside the test masses themselves; this is the so called mirror thermal noise. Here we present the status of the research concerning different mirror thermal noises, the attractive materials and we argue about the future research in the field.

The total thermal noise taken into account in this paper branches off three other noises: Brownian noise, thermo refractive noise and thermo elastic noise. It is important to note that the mirror thermal noise concerns both the substrate and the coatings. A good overview of the different thermal noises affecting the substrate and the coating of the interferometer mirrors can be found in reference [1]. On the basis of the formula given in the reference mentioned above, we have evaluated the thermal noise according to the parameters of different potential materials. All the expression given in the following of this paper consider a semi-infinite mirror, a Gaussian beam and an adiabatic frequency range (i.e. well below the mirror internal resonant frequency).

This report lists the thermal noise effect expected for different configuration of mirrors at room temperature and cryogenic temperature. Calculations use S.I. units. The symbols used are listed below:

- w Gaussian beam radius defined as the radius where the field amplitude is 1/e of the maximum (meter)
- λ laser wavelength (1064 nm or 1550 nm for silicon substrate)
- f measurement frequency (Hz)

- ρ density (kg/m³)
- C heat capacity (J/(K kg)
- κ thermal conductivity (W/(m K))
- σ poisson's ratio
- $\alpha$ thermal expansion coefficient (1/K)
- n refractive index
- d coating thickness (m)
- β=dn/dT thermo-optic coefficient (1/K)
- Φ loss angle
- Y Young modulus (Pa)
- Kb Boltzmann's constant, 1.39×10²³ J/K
- T temperature (300 K or 10 K)
- ω optical frequency (1/s)

### a. The Brownian noise

In general, the sensitivity of a gravitational wave detector such as Virgo or in the future, ET, is largely limited by the Brownian damping of payloads. The mirror thermal noise leads to a statistical deformation of the mirror surface which become one of the dominant source of noise in the frequency range comprises between 50 and 500 Hz. This noise is directly related to mechanical loss noted Φ.

#### i. The substrate Brownian noise

The Brownian noises are calculated from the direct approach according to Levin's theory [2]. By assuming a constant $\Phi_s(f) = \Phi_s$, the thermal noise described below uses the direct approach [3]. The power spectrum of the mirror displacement $S_B^s$ due to its internal thermal noise can be deduced by the following expression [1]:

$$S_B^s = \frac{2k_b T \Phi_s(f)(1-\sigma_s^2)}{\pi^{3/2} Y_s w f}. \tag{1.1}$$

#### ii. The coating Brownian noise

The spectral density of the coating Brownian noise is given by the following expression [1, 4]:

$$S_c = \frac{2k_b T \Phi_c(f)(1-\sigma_s^2)}{\pi^{3/2} Y_s w f}, \tag{2.1}$$

where $\Phi_c$ is the coating mechanical losses.

The coating Brownian noise is deduced from the same relation used to evaluate the substrate Brownian noise. Nevertheless, to evaluate the coating losses one has to take into account the anisotropic properties of the coating such as the Young's modulus and the Poisson ratio parallel and perpendicular to the coating layers. The loss angles of the coating layers parallel and perpendicular to the mirror surface permit to characterize the anisotropy of the coating. The expression of the coating loss $\Phi_c$ can be found in reference [5]:

$$\phi_c = \frac{(d_L + d_H)(1 - 2\sigma_s)}{\sqrt{\pi} w (1 - \sigma_s)(1 - \sigma_{para})}$$
$$\times \left[ \frac{Y_{perp}(1 - \sigma_{para}) - 2\sigma_{perp}^2 Y_{para}}{Y_{perp}^2 (1 + \sigma_s)(1 - 2\sigma_s)} Y_s \phi_{perp} + \frac{Y_{para} \sigma_{perp} (\phi_{para} - \phi_{perp})}{Y_{perp}} + \frac{Y_{para}(1 + \sigma_s)(1 - 2\sigma_s)\phi_{para}}{Y_s(1 + \sigma_{para})} \right],$$
(2.2)

where $d_L$ and $d_H$ are respectively the total thickness of the low index and high index materials thicknesses. Moreover, to estimate the total thickness of high and low index layers of the coating we use the following formulas:

$$d_L = \frac{\lambda_o}{n_L} z_L,$$
(2.3)

$$d_H = \frac{\lambda_o}{n_H} z_H,$$
(2.4)

where $z_L$ and $z_H$ are units of wavelength; for example 0.25 corresponds to the standard quarter wavelength multilayer coatings. The others quantities used in equation (2.2) are given by the expression below:

$$Y_{perp} = \frac{d_L + d_H}{\frac{1}{Y_L} d_L + \frac{1}{Y_H} d_H},$$
(2.5)

$$Y_{para} = \frac{Y_L d_L + Y_H d_H}{d_L + d_H},$$
(2.6)

$$\phi_{perp} = Y_{perp} \left( \frac{\frac{1}{Y_L} \phi_L d_L + \frac{1}{Y_H} \phi_H d_H}{d_L + d_H} \right),$$
(2.7)

$$\phi_{para} = \frac{1}{Y_{para}} \left( \frac{Y_L d_L \phi_L + Y_H d_H \phi_H}{d_L + d_H} \right),$$
(2.8)

$$\sigma_{perp} = \frac{\sigma_L Y_L d_L + \sigma_H Y_H d_H}{Y_L d_L + Y_H d_H} \ . \tag{2.9}$$

In order to evaluate $\sigma_{para}$ one has to solve the following equation [6]:

$$\sigma_{para}^2 \left(Y_{perp}.A\right) + \sigma_{para}\left(2\sigma_{perp}^2 Y_{para} A + Y_{perp} Y_{para}\right) + \left(2\sigma_{perp}^2 Y_{para} A - Y_{perp} A + Y_{para}^2 \sigma_{perp}^2\right) = 0, \tag{2.10}$$

with the constant A:

$$A = \left[\frac{\sigma_L Y_L d_L}{(1+\sigma_L)(1-2\sigma_L)} + \frac{\sigma_H Y_H d_H}{(1+\sigma_H)(1-2\sigma_H)}\right]\frac{1}{(d_L + d_H)} \ . \tag{2.11}$$

The equation of $\sigma_{para}$ has only one positive root which will be reported in the simulation.

If we consider a SiO$_2$ substrate and a coating of 17 doublets of Ti:Ta$_2$O$_5$-SiO$_2$, the evaluation of $\sigma_{para}$ gives the values of 0.2011. Reference [1] gives the following approximation:

$$\sigma_{para} = \frac{\sigma_L + \sigma_H}{2} = 0.1985 \ . \tag{2.12}$$

This is only 0.9% different respect to exact value. Nevertheless, in order to be as accurate as possible, we have used the exact formula instead of the approximation.

### b. The Thermo elastic noise

The thermo elastic deformation is a linear and/or volume dilatation or contraction under the influence of the temperature fluctuations. An alternative but equivalent way to see the same phenomena is explained in reference [7]: in response to the strain imposed, the material has tendency to heat or cool down according to the Le Châtelier principle. This creates a temperature gradient which dissipates a part of mechanical energy used to bend the material. Due to the fluctuation-dissipation theorem this dissipative process produces a fluctuation of the system. This thermal noise can limit the sensitivity of the detector and must be taken into account in evaluating the total thermal noise [8].

The thermo elastic noise has been proposed as a serious barrier limiting sensitivity of the gravitational wave detector [9]. The leading parameters are the thermal expansion coefficient $\alpha$ that generates fluctuations in the substrate (section I.b.i) but also in the coating (section I.b.ii).

### i. The substrate thermo elastic noise

The displacement of the mirror surface due to the substrate thermo elastic noise is given by the following expression suggested by Braginsky et al. [10]:

$$S_{TE}^s = \frac{4k_bT^2\alpha_s^2(1+\sigma_s)^2\kappa_s}{\pi^{5/2}(C_s\rho_s)^2 w^3 f^2}, \tag{3.1}$$

where $\alpha_s$ is the thermal expansion coefficient of the substrate, $\kappa_s$ is the thermal conductivity of the substrate, $C_s$ is the specific heat of the substrate and $\rho_s$ is the density of the substrate.

The limit of the adiabatic hypothesis, which is at the base of the expression (3.1) and which fails at low temperatures, has been explored in [11]. As it is explained in [11], the heat diffusion length increases considerably at low temperature and hence, becomes larger than the laser beam size, so that the adiabatic approximation is no longer valid. There is an angular frequency noted $\omega_c$ below which the adiabatic assumption fails. This frequency is a function of the material thermal conductivity, of the heat capacity and of the beam size:

$$\omega_c = \frac{4\kappa}{\rho C w^2}. \tag{3.2}$$

Following [11], the thermo-elastic noises can be written as:

$$S_{TE-Ad}^s = \frac{8k_bT^2\alpha_s(1+\sigma_s)^2 w}{\sqrt{2\pi\kappa}} J[\Omega], \tag{3.3}$$

with $\Omega = \omega/\omega_c$

and $\quad J[\Omega] = \sqrt{\frac{2}{\pi^3}} \int_0^\infty du \int_{-\infty}^{+\infty} dv \frac{u^3 e^{-u^2/2}}{(u^2+v^2)[(u^2+v^2)^2+\Omega^2]}. \tag{3.4}$

When $\omega \gg \omega_c$, the formula above coincides with (3.1). The expression has been taken into account in all the simulation at low temperatures since at this temperature the adiabatic approximation fails.

### ii. The coating thermo elastic noise

The thermo elastic noise of the coating has been described in several articles [1, 12, 13]. The model developed by Fejer et al. [13] includes a frequency dependent term $G(\omega)$:

$$S_{TE}^c = \frac{2k_bT^2}{\pi^{3/2}\sqrt{\kappa_s C_s\rho_s} w^2 f^{1/2}} d^2 \left(\alpha_c - \overline{\alpha}_s \frac{\langle C\rho\rangle}{C_s\rho_s}\right)^2 G(\omega), \tag{4.1}$$

where d is the total thickness of the coating and $\alpha_c$ and $\overline{\alpha}_s$ are the effective thermal expansion coefficients of the coating and the substrate defined as:

$$\overline{\alpha}_x = \alpha_x \frac{1+\sigma_s}{1-\sigma_x}\left[\frac{1+\sigma_x}{1+\sigma_s}+(1-2\sigma_s)\frac{Y_x}{Y_s}\right]$$

(4.2)

$$\alpha_c = \sum_{x=1}^{N} \overline{\alpha}_x \frac{d_x}{d}$$

$$\overline{\alpha}_s = 2\alpha_s(1+\sigma_s).$$

(4.3)

G(ω) is given by :

$$G(\omega) = \frac{2}{R\xi^2} \frac{\sinh\xi - \sin\xi + R(\cosh\xi - \cos\xi)}{\cosh\xi + \cos\xi + 2R\sinh\xi + R^2(\cosh\xi - \cos\xi)},$$

(4.4)

with:

$$\xi = \sqrt{2\omega d^2 \langle(\rho C)\rangle\left\langle\frac{1}{\kappa}\right\rangle}$$

$$R = \sqrt{\frac{\langle\rho C\rangle}{C_S \rho_S \kappa_S \left\langle\frac{1}{\kappa}\right\rangle}}$$

(4.5 and 4.6)

where N is the number of doublets in the coating.

Since the coating is formed by a stack of high index and low index thin films, it cannot be considered as a homogenous and uniform thin film. Fejer et al. [13] proposed to calculate the property of the multilayer coating by making the proper average of the properties of the two materials constituting it. This average enables to obtain a nice precision and suitable estimation of thermo elastic noise.

Consequently, all the terms represented under brackets have to be written according to the formula:

$$\langle X \rangle = \frac{X_L d_L + X_H d_H}{d}.$$

(4.7)

### c. The thermo refractive noise

The multilayer coatings are made with alternating low and high refractive index thin films noted respectively $n_L$ and $n_H$. Usually $SiO_2$ is the low index material and $Ta_2O_5$ is the high index material. Since $n_L$ and $n_H$ depend on the temperature T, the thermo optic coefficient $\beta$ = dn/dT is usually not zero. Consequently, a temperature fluctuation will induce a variation of the film optical thickness [10]. Similarly whenever a beam is transmitted through a substrate, a temperature fluctuation will induce a variation of the substrate optical thickness.

### i. The substrate thermo refractive noise

Currently, there are a few documents reporting the substrate thermo refractive noise. An estimate of the phase noise due to the thermo-refractive noise in the beam splitter substrate can be found in [14]:

$$S^s_{\Delta\varphi,TR}(f) = \left(\frac{\beta_s l}{\lambda}\right)^2 \frac{\sqrt{2}k_b T^2}{\pi w^2 \sqrt{\kappa_s \rho_s C_s 2\pi f}}, \tag{5.1}$$

where l is the thickness of the beam splitter. This phase noise is equivalent to a displacement noise of a test masses equal to:

$$\Delta x = \frac{\Delta\phi \cdot \lambda}{8F}, \tag{5.2}$$

where F is the Fabry-Perot cavity finesse that can be expressed from the amplitude reflectivities $r_1$ and $r_2$ of the two mirrors involved in the cavity as:

$$F = \frac{\pi\sqrt{r_1 r_2}}{1 - r_1 r_2}. \tag{5.3}$$

To calculate the influence of this noise, we have used the finesse proposed for Advanced Virgo, i.e., F = 885.

### ii. The coating thermo refractive noise

The coating thermo-refractive noise is given by the following expression [10]:

$$S^c_{TR} = \frac{2k_b T^2 \beta_c^2 \lambda^2}{\pi^{3/2} \sqrt{\kappa_s \rho_s C_s} w^2 f^{1/2}}, \tag{6.1}$$

where λ is the wavelength of the detector laser and $\beta_c$ is the effective coating thermo refractive coefficient defined by the following expression:

$$\beta_c = \frac{1}{4} \frac{dn_H/dT \, n_L^2 + dn_L/dT \, n_H^2}{n_H^2 - n_L^2}. \tag{6.2}$$

It is important to note that the equation is correct only for a coating made entirely of quarter wavelength doublets. In the case of a coating made of quarter wavelength doublets and ending with an half wavelength low index layer the proper equation is the following [16]:

$$\beta_c = \frac{1}{4} \frac{dn_H/dT \, n_L^2 + dn_L/dT \, n_H^2}{n_L^2(n_H^2 - n_L^2)} + \frac{dn_L/dT}{4n_L^2} \tag{6.3}$$

Both in (6.2) and (6.3) the coating thermal expansion coefficient has been neglected with respect to the dn/dT parameter. Following [16], in order to have the exact expression, one should replace $dn_x/dt$ with $dn_x/dT+\alpha_x \times n_x$.

### d. TE-TR compensation or the thermo optic noise

In March 2008, Kimble [15] came with the idea that thermo-elastic and thermo-refractive fluctuations should be correlated and that in some conditions can compensate each other thus reducing the total thermo-optic noise. On the one hand, the effect of a change in the index of refraction leads to change of the coating optical thickness and moves the effective mirror surface in one direction. On the other hand, a thermal expansion due to the same temperature fluctuation moves the surface of the mirror in a direction which depends on the relative sign of the thermal expansion coefficient and of the dn/dT parameter.

Since then, two authors [1, 16] have analysed and explained the problem following two slightly different approaches. Following [16] the total thermo-optic noise is given by the expression below:

$$S_{TER}^c = \frac{2k_b T^2}{\pi^{3/2} w^2 \sqrt{\kappa_s C_s \rho_s f}} \Gamma_{tc} \left( \alpha_c d - \beta_c \lambda - \overline{\alpha}_s d \frac{\langle C\rho \rangle}{C_s \rho_s} \right)^2, \quad (7.1)$$

where $\alpha_c$ and $\overline{\alpha}_s$ are effective expansion coefficient for a coating with N layers each of thickness $d_x$ and the effective expansion of the substrate, as defined in (4.1).

The coating specific heat $\langle C\rho \rangle$ is determined from a simple volume average:

$$\langle C\rho \rangle = \sum_{x=1}^{N} C_x \rho_x \frac{d_x}{d}. \quad (7.2)$$

The parameter $\Gamma_{tc}$ is a thick coating correction. This correction factor is given by the expression below:

$$\Gamma_{tc} = \frac{p_E^2 \Gamma_0 + p_E p_R \xi \Gamma_1 + p_R^2 \xi^2 \Gamma_2}{R \xi^2 \Gamma_3}, \quad (7.3)$$

with

$$\Gamma_0 = 2(\sinh(\xi) - \sin(\xi)) + 2R(\cosh(\xi) - \cos(\xi)), \quad (7.4)$$

$$\Gamma_1 = 8\sin(\xi/2)(R\cosh(\xi/2) + \sinh(\xi/2)), \quad (7.5)$$

$$\Gamma_2 = (1+R^2)\sinh(\xi) + (1-R^2)\sin(\xi) + 2R\cosh(\xi), \quad (7.6)$$

$$\Gamma_3 = (1+R^2)\cosh(\xi) + (1-R^2)\cos(\xi) + 2R\sinh(\xi). \quad (7.7)$$

Where $p_E$ and $p_R$ can be expressed as:

$$p_E = \frac{\left[\alpha_c - \overline{\alpha_s}\frac{\langle C\rho \rangle}{C_s\rho_s}\right]d}{\left[\alpha_c - \overline{\alpha_s}\frac{\langle C\rho \rangle}{C_s\rho_s}\right]d - \beta_c\lambda} \quad (7.8) \quad \text{and} \quad p_R = \frac{-\beta_c\lambda}{\left[\alpha_c - \overline{\alpha_s}\frac{\langle C\rho \rangle}{C_s\rho_s}\right]d - \beta_c\lambda} \quad (7.9)$$

The compensation has been taken in account in all simulations of this document.

### e. The total thermal noise

The total thermal noise has been evaluated according to a sum of the individual noises. More precisely, the thermal noise must be equal to the square root of the sum of individual noise. Since the coating and substrate Brownian noise, are two independent noises, the total Brownian noise is given by the sum of the square noise. On the contrary, the coating thermo refractive and thermo elastic noises are dependent from each other [1] as they both depends on the same temperature fluctuation. Thus, the total thermo elastic/refractive noise is given by the linear sum of the noises. In conclusion, the total thermal noise has been evaluated using following equation:

$$S_{TN} = \sqrt{S_B^c + S_B^s + S_{TE}^s + S_{TR}^s + S_{TER}^c} \, . \quad (8.1)$$

## 2. Materials properties at room temperature

### a. Substrates

In this note three materials have been considered for the mirror substrates: fused silica, sapphire and silicon. The thermo-mechanical and optical properties of these materials that are relevant for thermal noise are given in Table 1.

#### i. Silica

Currently, all the gravitational wave detectors working at room temperature use fused silica (LIGO, VIRGO, GEO600) as mirror substrates. The main reason for this choice has been the excellent optical quality of this material and the possibility to obtain very high quality surface by means of the appropriate polishing method.

In addition, fused silica has quite low intrinsic mechanical losses. At the time LIGO and VIRGO were built values as low as $5\times10^{-8}$ [9] were known. Since then it has been shown that the mechanical losses of Suprasil (the kind of silica used for the most critical optical components) depends upon the frequency considered and the surface/volume ratio. The measurements of the mechanical loss in fused silica have been compiled from samples having different geometries and resonant frequencies in order to model the known variation of the loss with frequency and surface to volume ratio. With this model, mechanical losses of fused silica at different frequency and substrate dimension can be predicted. For example, the loss angle prediction for the VIRGO mirror substrates is $5\times10^{-9}$ [12]. This value is reported in table 1 and has been used in the simulation. Another remarkable property of fused silica is its low expansion coefficient. As a consequence fused silica has relatively low thermo-elastic noise.

### ii. Sapphire

Sapphire is a single crystal of aluminium oxide ($Al_2O_3$). The crystal has a hexagonal crystalline structure that shows evidence of anisotropy in many optical and physical properties. Therefore, the orientation of the optical axis or c-axis can cause some of the relevant material properties to be different.

A Sapphire substrate offers key advantages for cavity mirrors of gravitational wave detectors. First, its optical properties show a perfect transparency at 1064 nm and a low dn/dT. Regarding of thermal properties, sapphire offers excellent heat dissipation thanks to its good thermal conductivity important for the cooling of the mirrors. However, polishing is more difficult on sapphire; in particular, achieving good polish was unsolved in 2004 [17].

The use of sapphire as mirror substrate has been considered for Advanced LIGO. Indeed, sapphire presents the advantage to have low mechanical losses and high Young's modulus. However, the choice of sapphire at room temperature has been finally turned down because of its large thermal expansion coefficient ($5.1\times10^{-6}$) which is about 10 times larger than that of fused silica [18]. At low temperatures, sapphire shows several benefits: good optical properties, high thermal conductivity, small temperature coefficient of the refractive index [19].

The properties of sapphire as substrate material have been intensively studied by the Japanese groups [20]. At room temperature the mechanical losses were found to be $2.17\times10^{-7}$. This value is in disagreement with the data from [12] ($3\times10^{-9}$). Nevertheless this result was limited by the surface roughness. Sapphire substrate has been, indeed, possibly damaged due to machining and polishing that generates small crystallites of hundred microns deep. For this reason the value used in the simulation for the mechanical loss of the sapphire substrate is $2\times10^{-9}$.

Cooled sapphire payloads are being studied in Japan as part of a proposal to construct the large-scale cryogenic gravitational wave telescope (LCGT) [21] but some analysis suggests only moderate performance at low temperature for sapphire [22].

### iii. Silicon

Silicon substrates are an alternative to sapphire at low temperature. Silicon has the advantage of a thermal expansion coefficient that crosses zero [23] around 18 K and 123 K. Nevertheless, the other optical and thermal properties of silicon and sapphire are comparable. The simulation in the second part of this note will permit to discuss the comparison.

As in the case of sapphire also in silicon the crystal orientation can change the properties of material depending on the direction considered and moreover, silicon has a doping effect (p or n) that begins to be studied and that can changed some values as mechanical losses [24]. If silicon proves to be a good candidate as substrate for GWD, the wavelength of the optical system must be changed from 1064 nm to a value in the range 1400-1600 nm. This leads to a thicker coating and hence, a more important source of noise. An alternative could be all-reflective topologies currently under investigation. In table 1 both the values of the index of refraction at 1064 and 1550 nm are given. In the numerical simulation the wavelength used in the case of silicon is supposed to be 1550 nm.

### iv. Parameters values

| | SiO$_2$ | Silicon | Sapphire |
|---|---|---|---|
| Loss angle | **5×10$^{-9}$** [1, 12, 25]<br><br>5×10$^{-8}$ [9, 10] | 3×10$^{-8}$ [26, 27],<br><br>**1×10$^{-8}$** [28]<br><br>3×10$^{-9}$ [29] | 3×10$^{-9}$ [12, 25]<br><br>2.17×10$^{-7}$ [20]<br><br>**2×10$^{-9}$** [30] |
| Density (kg.m$^{-3}$) | 2200 [1, 12] | 2330 [31] | 3980 [7] |
| Thermal conductivity (W.m$^{-1}$.K$^{-1}$) | **1.38** [1] | 130-160 [31] (Ave. **145**) | 36 [2, 7] – 40 [10, 12]<br><br>**33** [13], 46 [32, 33] |
| Specific heat (J.K$^{-1}$.Kg$^{-1}$) | **746** [1], 670 [10, 12] | **711** [31] | **770** [7], 790 [10, 12] |
| Thermal expansion coef. (K$^{-1}$) | 0.55×10$^{-6}$ [10, 12, 34]<br><br>**0.51×10$^{-6}$** [1, 35] | **2.54×10$^{-6}$** [31] | **5.1×10$^{-6}$** [7, 10, 12]<br><br>5.4×10$^{-6}$ [13]<br><br>6.6×10$^{-6}$ [36] |
| Thermo optic coef.: dn/dT | 1.5×10$^{-5}$ [1, 10]<br><br>**8×10$^{-6}$** [37, 38] | 1.87×10$^{-4}$ [39] [40]<br><br>**5.15×10$^{-5}$** [41] | **1.3×10$^{-5}$** [33] [42] |
| Young's modulus (GPa) | 72 [1, 10] | **162.4** [31]<br><br>(100)130, (110)169, (111)188 [43] | 400 [10, 12, 13, 36] |
| Poisson's ratio | 0.17 [1, 10, 12] | 0.22-0.28 [44], **0.22** [45] | 0.23$^7$ -0.29 [10, 12]<br><br>**0.235** [45] |
| Refractive index @ 1064 nm | 1.45 [1, 12] | 3.543 [39] @ 1064 nm<br><br>**3.453** [39] @ 1550 nm | 1.75 [32] |

Table 1. List of the values of different material substrate parameters at 300 K. The value in bold are those used in the numerical simulations

### b. Coatings

In table 2 are given the thermo-mechanical and optical parameters for several optical coatings at room temperature.

| | SiO$_2$ | TiTa$_2$O$_5$ | Ta$_2$O$_5$ | TiO$_2$ | Al$_2$O$_3$ | Nb$_2$O$_5$ | ZrO$_2$ | HfO$_2$ |
|---|---|---|---|---|---|---|---|---|
| Loss angle | 0.4×10$^{-4}$ [37]<br>**0.5×10$^{-4}$** [46]<br>10$^{-3}$ on sapphire [47] | 2.3×10$^{-4}$ [37]<br>**2×10$^{-4}$** [36] | 3.8×10$^{-4}$ [1] | 6.3×10$^{-3}$ deduced from [48] | **2.4×10$^{-4}$** on silica [47]<br>2×10$^{-5}$ on sapphire [47] | 6.7×10$^{-4}$ [47]<br>**4.6×10$^{-4}$** [49] | 2.3×10$^{-4}$ [36] | 5.9×10$^{-4}$ [50] |
| Density (kg.m$^{-3}$) | 2200 [1, 12] | 6425 [36] | 6850 [1] | 4230 [51] | 3700 [52] | 4470 [53]<br>**4590** [45] | **6000** [54]<br>5750 [55]<br>6100 [45] | 6400-**8000** [56]<br>9674 [45] |
| Thermal conductivity (W.m$^{-1}$.K$^{-1}$) | 1,38 [1, 37]<br>0.2-**0.5** [57] | 33 [13]<br>0.2-**0.6** (estimated) | 33 [13]<br>0.3-**0.6** [57]<br>0.35 [58] | 1.15 [59]<br>**0.45** [58] | **3.3** [14]<br>0.7-1.1 [57] | Assumed quite low as : 1 [60] | **1.09** [23]<br>2 [54], 1.94 [45]<br>0.2 [58] | 0.07-0.2 [57]<br>**1.2** [61]<br>2.36 [45] |
| Specific heat (J.K$^{-1}$.Kg$^{-1}$) | **746** [1]<br>670 [10] | 269 [37] | **306** [13]<br>174 [62] | **130** [59]<br>40 [62] | 310 [63] | 590 [64]<br>499 [45] | **26** [61]<br>24 [55] | 16.7 [61] |
| Thermal expansion coef. (K$^{-1}$) | 0.51×10$^{-6}$ [1, 35, 37] | 3.6×10$^{-6}$ [37] | **3.6×10$^{-6}$** [1, 14, 65]<br>5×10$^{-6}$ [14] | 5×10$^{-5}$ [12] | 8.4×10$^{-6}$ [52] | 5.8×10$^{-6}$ [65] | 8.8×10$^{-6}$ [62]<br>**10.3×10$^{-6}$** [54, 55] | 3.8×10$^{-6}$ [23] |
| Thermo optic coef.: dn/dT | 1.5.10$^{-5}$ [1]<br>**8×10$^{-6}$** [37, 38] | 14×10$^{-6}$ [37] | 1.2×10$^{-4}$ [12],<br>6×10$^{-5}$ [1]<br>**2.3×10$^{-6}$** [66] | -1.8×10$^{-4}$ [12] | 1.3×10$^{-5}$ [33] | 1.43×10$^{-5}$ [65] | 3-60×10$^{-5}$ [67]<br>**10x10$^{-5}$** for simulation | |
| Young's modulus (GPa) | 72 [1, 10, 37]<br>40-**60** [14] | 140 [37] | 140 [37] | 290 [12] | 210 [68] | 60 [65]<br>**68**-102 [53] | 200 [54] | 380 |
| Poisson's ratio | 0.17 [1, 10] | 0.23 [6] | 0.23 [37] | 0.28 [12] | 0.22 [52] | 0.2 [65] | 0.23-0.31 [55]<br>**0.27** [45] | *0.2 assumed* |
| Refractive index @ 1064 nm | 1.45 [1, 12] | 2.065 [1],<br>**2.06** [37],<br>2.07 [36],<br>2.03 [13] | 2.06 [37]<br>2.03 [36] | 2.3 [12] | 1.63 [5] | 2.32 [65]<br>**2.21** [49] | 2.15 [36, 69]<br>**2.1** [36] | 2.08 [69] |

Table 2. List of the values of different coating materials parameters at 300 K.

Both Tables 1 and Table 2 show an important variation of the parameters according to different sources. The data require vigilance and extensive research. Several points must be pointed out:

In the past, the thermal conductivity of $Ta_2O_5$ has been assumed to be closer to that of sapphire [13]. However, the article [57], reports the thermal conductivities of several thin films materials like $SiO_2$, $Al_2O_3$, $Ta_2O_5$ and $HfO_2$. The thermal conductivities from [57] detailed in the table 2, have been measured for a single-layer of coating by different groups and by several photo thermal methods. The difference between the values depends on the photo thermal method used (radiometry, surface displacement, mirage, thermal pulse delay) and the thickness of the thin film. Another article from 1993 confirms the same order of magnitude for the thermal conductivity of different thin oxide films ($TiO_2$, $Ta_2O_5$, $ZrO_2$) [58].

Article [57] shows that thermal conductivity values of the analyzed thin films are significantly lower than those of the corresponding materials in bulk form. For example, $HfO_2$ bulk is around 17 $Wm^{-1}K^{-1}$ whereas the thin film conductivity is less than 0.2 $Wm^{-1}K^{-1}$. Sapphire substrates have a thermal conductivity of 33 W/m K and the parameters value decrease at 3.3 W/m K for $Al_2O_3$ thin film. Furthermore, the thermal conductivity data are strongly dependent on impurities especially at temperature below 300 K [45]. Thermal expansion coefficient and Young's modulus can also change between bulk and thin film for a same material. For instance, the thermal expansion of $SiO_2$ thin films have been observed in the range of $0.6$-$4\times10^{-6}$ with a Young's modulus of 40-60 GPa whereas $SiO_2$ bulk has a thermal expansion of $0.5\times10^{-6}$ and a Young's modulus of 72 GPa [14].

Some parameters have been estimated. The Poisson ratio of $HfO_2$ has been assumed as 0.2 because most materials have between 0 and 0.5. The approximation takes also into account the value of other Poisson Ratio of different oxide thin film.

Another important detail is the difference for a deposited thin film mechanical losses according to the substrate [47]. While a thin film of $Al_2O_3$ deposited on silica has a loss angle of $2.4\times10^{-4}$, deposited on sapphire the loss angle decreases to $2\times10^{-5}$. This particularity deserves more investigations.

### 3. Mirrors materials properties at cryogenic temperature

The sensitivity of the GWD is limited by thermal noise of the payload (mirrors and suspensions) within the detection band of about 50-500Hz. Cooling down the payload of the gravitational waves detector to cryogenic temperature is a promising technique to reduce the thermal noise and consequently increase the sensitivity of the detector. While there is a great interest of published data on cryogenic properties, it is often difficult to find the information about the most critical parameters because the data are mostly dispersed in many publication, books, notes, etc. This part of the paper will be focused on substrate and coating materials properties at low temperature. It is meant to be a starting point to compare different solutions to realise a cryogenic mirror both in terms of substrates and coatings.

### a. Substrates

| | SiO$_2$ | Silicon | Sapphire |
|---|---|---|---|
| Loss angle | 10$^{-3}$ [43] | 1.1×10$^{-8}$ @10 K [27]<br>10$^{-9}$ @10 K [24] | 4×10$^{-9}$ @ 4.2 K [20] |
| Density (kg.m$^{-3}$) | 2220 @ 80K [45] | 2331 @ 10 K [45] | 3997 @ 20 K [45] |
| Thermal conductivity (W.m$^{-1}$.K$^{-1}$) | 0.25 @ 4,2K and 0.7 @ 20K [51]<br>**0.4 @ 10K** | **2330 @ 10 K** [45]<br>297 @ 4 K [45]<br>598.6 @ 125 K [70] | 110 @ 4.2 K [51, 61]<br>**1500 @ 12.5 K** [61]<br>4300 @ 20 K [33] |
| Specific heat (J.K$^{-1}$.Kg$^{-1}$) | 1-7 @ 10 K [63]<br>**3 @ 10 K** [11]<br>0.26 @ 20 K [45] | 0.276 @ 10 K [45] | 9.49 @ 90 K [71]<br>**9.34×10$^{-2}$ @ 10 K** [45] |
| Thermal expansion coef.(K$^{-1}$) | -0.25×10$^{-6}$ @ 10 K [63] [11] | 4.85×10$^{-10}$ @ 10 K [45] | 5.3×10$^{-10}$ @ 10 K [71] |
| Thermo optic coef.: dn/dT | 1.01×10$^{-6}$ @ 30 K [72] | 5.8×10$^{-6}$ @ 30 K and 1550 nm [39] | **9×10$^{-8}$** @ 4 K [33]<br>from 5K to 40 K :<br>-6×10$^{-8}$ <β< 9×10$^{-8}$ [19] |
| Young's modulus (GPa) | 71.2 @ 10 K [45] | 162.4 | 464 @ 10 K [45] |
| Poisson's ratio | 0.159 @ 10 K [63] | 0.2205 [45] | *0.23* |
| Refractive index @ 1064 nm | 1.44876 @ 30K [72] | 3.45 @ 30 K and 1550 nm [39] | 1.75 [32] |

Table 3. List of the values of different material substrate parameters at low temperature.

Table 3 lists the cryogenic properties of potential substrate materials that have been used in the simulations. The cryogenic temperatures have an impact on some parameters, notably; the thermal conductivity that tends to decrease at low temperature and exhibits a peak for higher temperature (between 20 to 80 K) which depends on the materials. As a consequence there is an important difference between the thermal conductivity at 4 K and 10 K. The values at 10 K have been used in the simulation since it is expected that the temperature of a cryogenic mirror will be around this value. When the data at 10 K are not available; the nearest known data have been used.

### b. Coatings

|  | SiO$_2$ | Ta$_2$O$_5$ | TiO$_2$ | Al$_2$O$_3$ | HfO$_2$ |
|---|---|---|---|---|---|
| Loss angle | 6×10$^{-4}$ [43]<br><br>5×10$^{-4}$ | 3.8×10$^{-4}$ (doped) [50] | 5.6×10$^{-3}$ deduced from [48] @77K |  | 2.2×10$^{-4}$ [50] |
| Density (kg.m$^{-3}$) | 2200 [1, 12] | 6850 [1] | 4230 [51]<br><br>4269 @ 73 K [45] | 3700 |  |
| Thermal conductivity (W.m$^{-1}$.K$^{-1}$) | 0.13 @10K [63] |  | 500 @ 10 K [51]<br><br>1500 @ 10 K [61] | 2-5 @ 10 K [63]<br><br>5 @ 10 K [61] |  |
| Specific heat (J.K$^{-1}$.Kg$^{-1}$) | 1-7 @ 10 K [63]<br><br>(4) | 3.17 @ 50 K [62] | 0.012 @ 10 K [62] | 0.1 @ 10 K [63]<br><br>5×10$^{-3}$ @ 10K [62] |  |
| Thermal expansion coef. (K$^{-1}$) | -0.25×10$^{-6}$ @ 10 K |  | 6.5×10$^{-6}$ @ 100 K [23, 45] | 0.6 ×10$^{-6}$ @ 100 K [23] |  |
| Thermo optic coef.: dn/dT | 1.01×10$^{-6}$ @ 30 K [73] |  |  |  |  |
| Young's modulus (GPa) | 60 | 140 | 290 | 356 |  |
| Poisson's ratio | 0.159 @ 10 K [63] | 0.21 | 0.25 | 0.2 |  |
| Refractive index @ 1064 nm | 1.44876 @ 30 K [72] | 2.05 | 2.28 | 1.61 |  |

Table 4. List of the values of different coating materials parameters at 10K.

Although, the mechanical loss of Ti:Ta$_2$O$_5$ has been intensively studied in the last years, there are no data concerning its characteristics in terms of thermal conductivity, specific heat, thermal expansion coefficient and thermo optic coefficient. First because it is a doped material not referenced and secondly because the values might change depending on the amount of doping. A campaign of characterisation is needed to make progress in this field.

Mechanical losses of coating materials are one of the major limitation to the sensitivity of GWD at room temperature [46]. Thus the behaviour of these mechanical losses at low temperature is a major question. Of course, the reduction of the temperature from 300 K down to 10 K should reduce by more than a factor of five the Brownian coating noise but, as it is shown in the table 3 and 4, several of the considered material parameters exhibit large variations at cryogenic temperatures. Therefore, it is important to evaluate other thermal noises depending on these parameters as the thermo elastic and the thermo refractive noise.

The doping of Ta$_2$O$_5$ with TiO$_2$ provides reduction in the mechanical dissipation at room and cryogenic temperatures. It is a matter of urgency to discuss the effect on the other thermal noises.

Low temperatures have also a serious impact on the specific heat, the thermal expansion and thermo optic coefficient. On the other hand other material parameters do not show significant changes; these include the density, the Young's modulus, the Poisson's ratio and the refractive index. In the absence of data these parameters will be considered as having the same values at room and at cryogenic temperature. Consequently, refractive index, Poisson's ratio, density and Young's modulus remain unchanged.

4. Simulation of the thermal noise

   a. Mirror thermal noise at room temperature

In order to compare the behaviour of different substrate materials we have calculated the thermal noise of a high reflectivity mirror made with different substrates and the same 'standard' coating. The 'standard' coating is a multilayer $(HL)_{17}$ HLL coating made of Ti:Ta$_2$O$_5$ and SiO$_2$ quarter wavelength layers. On silica and sapphire substrates, it corresponds to a transmission of 6 ppm. In the case of Silicon the working wavelength is supposed to be 1550 nm while for the other kind of substrates the laser wavelength is supposed to be 1064 nm. To have the same transmission, the coating is a multilayer $(HL)_{19}$ HLL on silicon substrate. As a consequence the coating layers considered for silicon are about 50% thicker. In the calculation the laser beam radius is equal to 6 cm which corresponds to an advanced detector beam. The best mechanical loss angles have been considered both for the substrates and the coating materials. The result of the simulation is shown in Figure 1: the dashed lines show various coating thermal noises while the continuous lines refer to the thermal noises in the substrate. For high reflective optics, the thermo-refractive noise of the substrate is negligible.

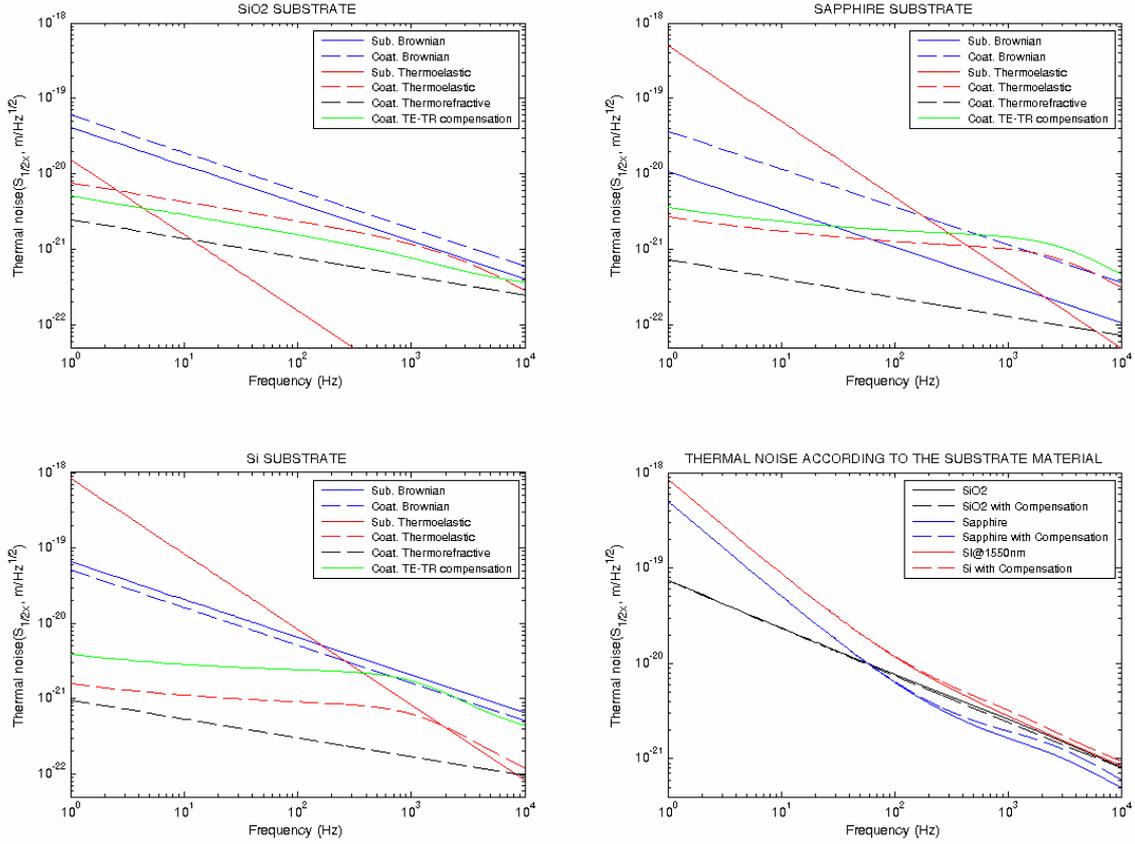

Figure 1. Evaluation of the thermal noises for different substrates at 300 K.

For the silica substrate, the coating Brownian noise is the limiting factor at all frequencies. The simulation shows the importance of developing better coatings for new generation of GWD and for Advanced Virgo.

In the case of a mirror using a sapphire substrate, the coating Brownian noise is the main limitation at high frequency while at low frequency; the substrate thermo elastic noise becomes the limiting factor.

Finally, at 300 K, silicon is unsuitable due to its large mechanical dissipation [26] [27] [28]. Figure 1 shows that the substrate Brownian noise is the limiting factor for silicon at high frequency while its performance is limited by the substrate thermo-elastic noise at low frequencies.

Silicon measurements reported in the three references where made on thin and small substrates. Nevertheless, a lower mechanical loss of 3 $10^{-9}$ were recently reached at room temperature for larger substrates [29]. However, for thermal noise calculation in silicon substrate, the value at $10^{-8}$ reported in [28] have been used. With a lower value, the Coating Brownian noise would be reduced allowing an reduction of the total thermal noise for high frequencies but would not change the silicon as a competitive substrate at room temperature compared to silica.

The green curve represents the compensation of the thermo-elastic and thermo-refractive noises described by M. Evans [16]. The compensation depends on substrate properties ($\alpha$, $\sigma$, C and $\rho$). For the silica substrate, the compensation improves the sensitivity. On the contrary, for silicon and sapphire substrates the sensitivity tends to increase. The compensation has a small influence on the total thermal noise at high frequencies. Nevertheless, the detector will be limited by photon shot noise in this frequency region.

For Advanced LIGO the sapphire was turned down because of its thermo elastic noise. As a consequence, the most adapted substrate for detectors like Advanced Virgo and Advanced LIGO that will operate at 300 K is fused silica. In order to compare the performances of different coatings a simulation was performed using different coatings having approximately the same optical response (between 3 and 4.3 ppm @ 1064 nm).

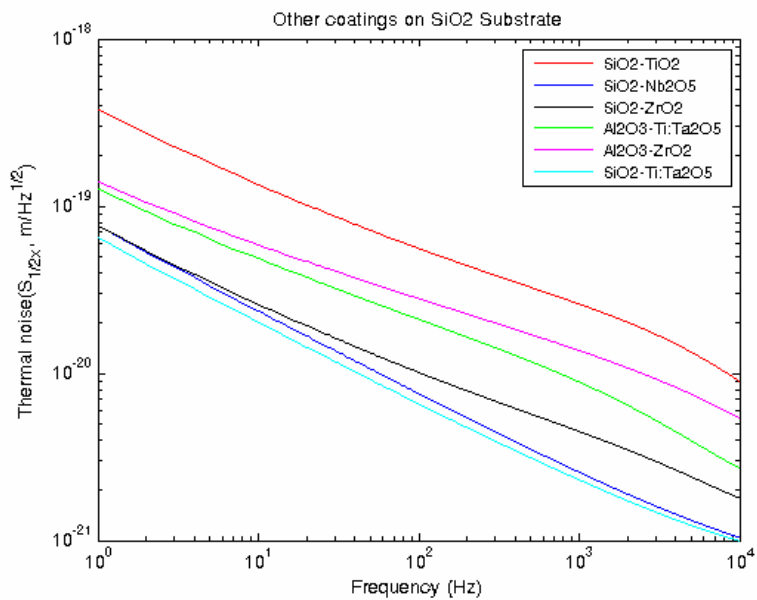

Figure 2. Evaluation of thermal noises for different coatings @ 300 K.

In order to have the same transmission different coating thickness have been simulated depending on the material used: $SiO_2$:$TiTa_2O_5$ have a ($(HL)_{18}$ HLL) configuration, $SiO_2$-$TiO_2$ ($(HL)_{14}$ HLL) configuration, $SiO_2$-$Nb_2O_5$ ($(HL)_{15}$ HLL), $SiO_2$-$ZrO_2$ ($(HL)_{17}$ HLL), $Al_2O_3$-$Ta_2O_5$ ($(HL)_{27}$ HLL) and $Al_2O_3$-$ZrO_2$ ($(HL)_{25}$ HLL). All the thermal noises for different coatings have been simulated on a $SiO_2$ substrate which has been qualified as the best substrate at room temperature. Figure 2 shows a clear advantage for the $SiO_2$-Ti:$Ta_2O_5$ coating. The results obtained for $SiO_2$-$Nb_2O_5$ and $SiO_2$-$ZrO_2$ are encouraging as well. Nevertheless, this work does not take into account the optic aspect (i.e. the optical absorption in the coating). The $SiO_2$-$TiO_2$ coating is the less attractive coating due to its high negative thermo optic coefficient and thermal expansion coefficient. The total thermal noises of these different coatings have been estimated taking into account the correlation of coating thermo elastic and thermo-refractive noise.

### b. Mirror thermal noise at cryogenic temperature

#### i. Behavior of substrates at low temperature

As explained in table 4, most of the coating parameters at cryogenic temperatures are missing, but all parameters have been found and referenced for the three substrate materials considered. In this part, we propose to explore the efficiency and the behaviour of substrates at different cryogenic temperatures (silicon substrate has been quoted as an example). The silica substrate is unsuitable at low temperature due to its high mechanical loss angle. As for them, silicon and sapphire substrates are two competitive materials that we propose to study.

First, a total thermal noise study for the two promising substrates has been done. Only the thermo elastic noise and the Brownian noise have been considered for these calculations. The thermo-refractive noise can be ignored. Its contribution does not change the total thermal noise within the frequency range of interest.

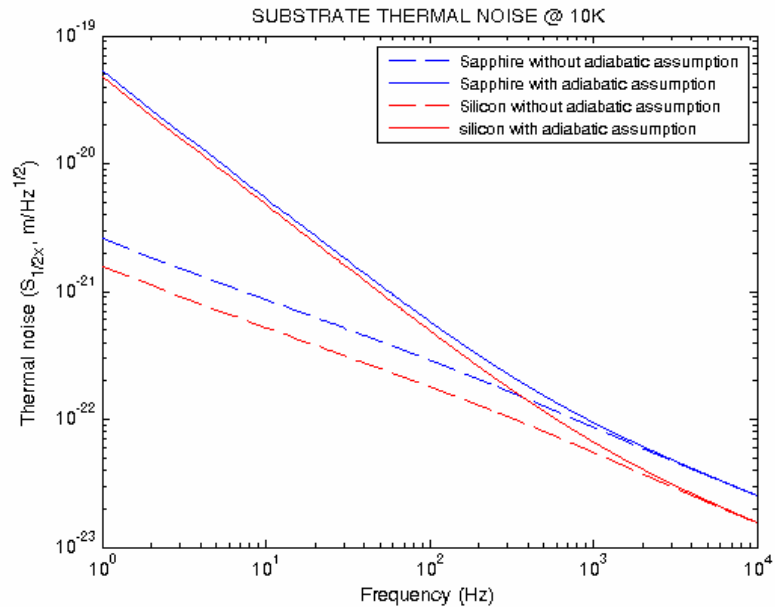

Figure 3. Evaluation of substrate thermal noise with and without adiabatic limit.

Figure 3 shows the evolution of total thermal noises on silicon and sapphire with and without the adiabatic assumption. The dashed curves use the standard equation referred by [10], it correspond to the adiabatic approximation. The continuous lines, on contrary, do not consider the adiabatic approximation.

The calculation has been done following [11]. In general, the noise at low frequencies is smaller than the one which would be obtained using the adiabatic approximation. Moreover, if $\omega \gg \omega_c$, the two spectral densities become the same. The results in figure 3 show that from the point of view of substrate thermal noise silicon is slightly better at cryogenic temperatures.

Then, to show the large variation of substrate parameters at low temperatures, the parameters of silicon and sapphire materials have been plotted and moreover, the total thermal noise of a silicon substrate at 4 K, 10 K, 20 K and 30 K has been simulated.

Figure 4 shows the thermal expansion coefficient, the thermal conductivity and the specific heat of silicon and sapphire as a function of temperature. The parameters are plotted from 0 K to 20 K. Specific heat and thermal conductivity have the same behaviour for both, silicon and sapphire substrates. They tend to increase with temperature. However, these parameters are larger in the case of silicon. The thermal expansion coefficient $\alpha$ at low temperatures varies from case to case : for sapphire, the $\alpha$ simply increases with temperature and for silicon, the $\alpha$ goes from positive to negative values crossing to zero around 18 K and 124 K. Consequently, as explained in [22], the thermo elastic noise of silicon should decrease with temperature and goes to zero at some specific temperatures.

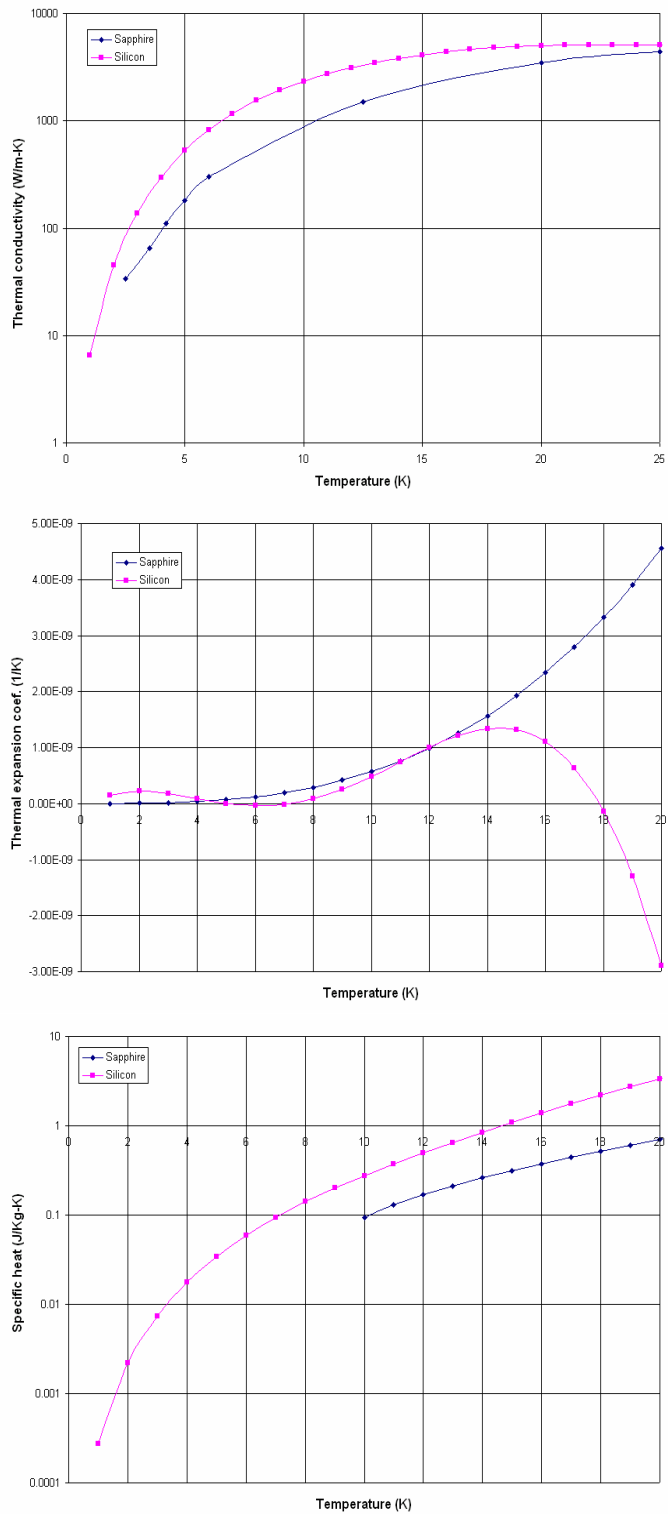

Figure 4. Evolution of silicon and sapphire thermal parameters at low temperature.

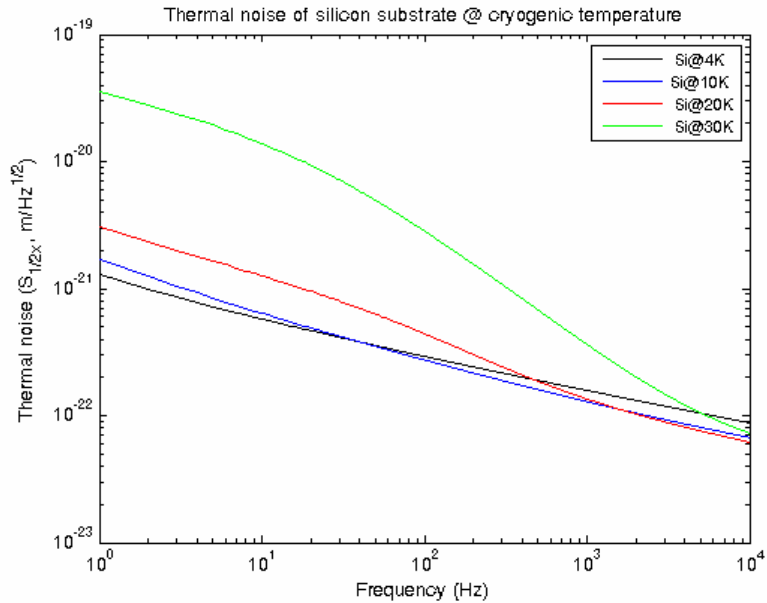

Figure 5. Evolution of the Silicon substrate thermal noise at different cryogenic temperatures.

Figure 5 shows the evolution of the Silicon substrate thermal noise at different cryogenic temperatures. At 4 K and 10 K the total thermal noise of the silicon substrate does not vary. The curves at 20 K and 30 K show the large variation of the thermal properties of silicon. As an example at low frequencies, from 20 to 30 K, the total thermal noise of the substrate increases by a factor 10.

On the basis of these results it is difficult to make a choice between sapphire and silicon substrates: The parameters at low temperatures follow the same behaviour from 0 to 20 K except for thermal expansion coefficient and the total thermal noises of silicon and sapphire are in the same order of magnitude. The optical point of view should be decisive.

### ii. High reflective mirror simulation at low temperature

In order to compare the behaviour of different substrate materials at low temperature we have calculated the thermal noise of a high reflectivity mirror made of different substrates and the same 'standard' coating. The 'standard' coating is a multilayer of $Ta_2O_5$ and $SiO_2$ $(HL)_{17}$ HLL for silica and sapphire substrates and $(HL)_{19}$ HLL for silicon substrates. The beam diameter is 6 cm. Some parameters of $Ta_2O_5$ are unknown at 10 K and have been consequently either estimated or taken at room temperature:

- Specific heat has been found at 50 K only,

- Thermal conductivity has been deduced from $TiO_2$ (nevertheless this parameter does not change the final simulation),

- For the thermal expansion and thermo-optic coefficients the room temperature values have been used.

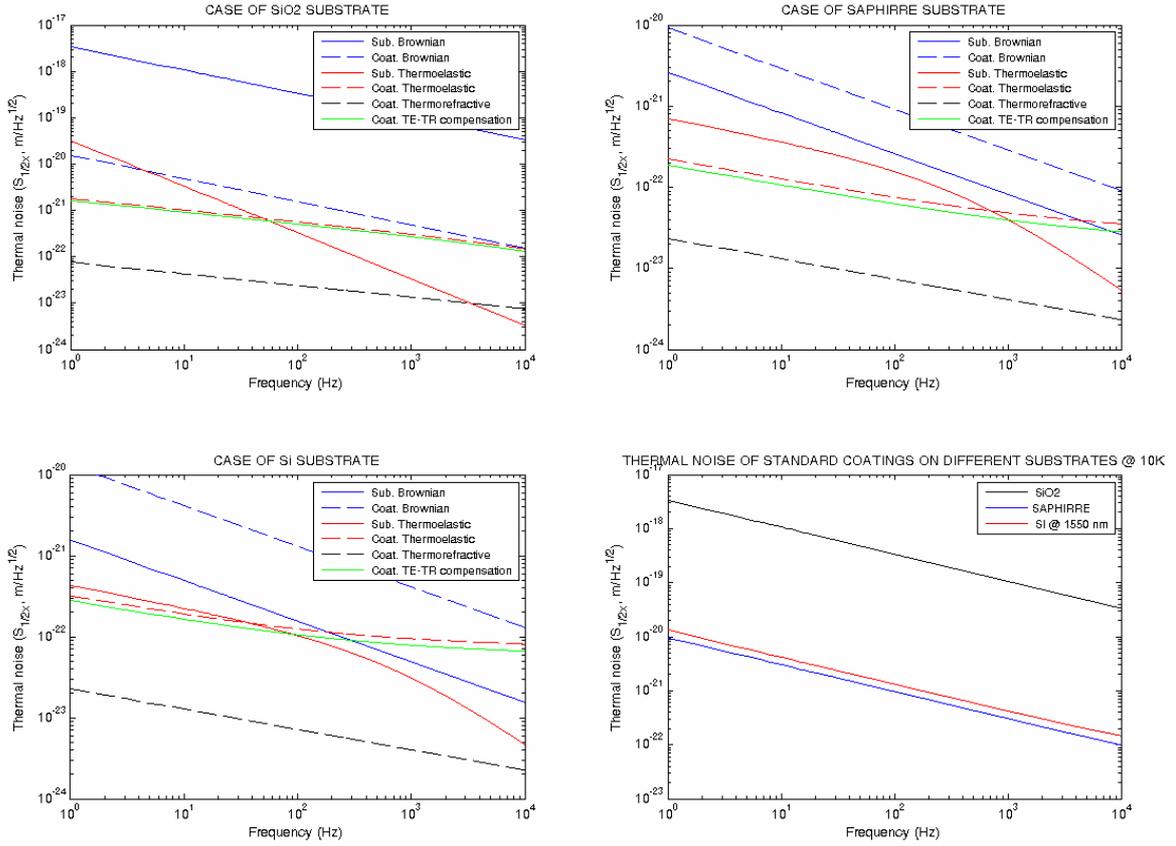

Figure 6. Evaluation of Total Thermal Noises @ 10 K for Ti:Ta$_2$O$_5$-SiO$_2$ coatings on different substrate materials: silica, sapphire, silicon.

For the silica substrate, the substrate Brownian noise is the limiting factor at all frequencies due to its large mechanical losses at low temperatures [43].

In the case of a mirror using a sapphire and silicon substrate, the coating Brownian noise is the main limitation at all frequencies. Due to its higher Young's modulus, sapphire gives slightly better results in a situation where the coating Brownian noise is the limiting factor. The simulation shows the importance of developing better coatings and of understanding the behaviour of coatings at low temperature for third generation of GWD.

## 5. Perspectives

Further improvements are possible in the next future, trying to achieve better thermal noise sensitivity.

Concerning coatings, the priority is to continue the study of the Ti:Ta$_2$O$_5$ properties at cryogenic temperatures. A detailed investigation and understanding of the loss mechanisms at cryogenic temperatures is necessary to improve these coatings. We have to entirely exploit also the existing and promising materials coating. In the next few years, testing the role of doped materials at cryogenic temperatures and other materials ($ZrO_2$, $Nb_2O_5$, $HfO_2$, etc) should be proceeding. For example, $HfO_2$ is an interesting material which shows low mechanical dissipation at low temperature. Added to this, it is necessary to explore other important and decisive parameters as specific heat, thermal conductivity, thermal expansion and thermo-optic coefficient which are usually missing.

Investigation from the optic point of view of coatings working at a 1550 nm should be also within the focus.

Another important task in the near future is also to compare silicon and sapphire from the optical point of view taking into consideration that these two substrates will need to work at two different wavelengths. The thermal noise aspect is currently not decisive, and the optic point of view must be taken in account.

## 6. Conclusion

We have evaluated the total mirror thermal noise at room and cryogenic temperatures by implementing a model that includes Brownian noise, thermo elastic and thermo refractive noise. The model used with the parameters listed above allows improving the understanding of the thermal noise at cryogenic temperature and shows that sapphire and silicon are two good test mass materials for the 3$^{rd}$ generation of GWD. At low temperatures, substrate thermal noise is slightly better for silicon while sapphire gives better results when coating Brownian noise is a limiting factor. The optical point of view will be an important element to make a decision. In any case coating Brownian noise is the limiting factor of currently available reflective components. The capability to improve its performances at 1550 nm (silicon) or 1064 nm (sapphire) could be another element for a decision which material to use. At room temperature silica remains the best available solution. When moving from 300 K to 10 K, the thermal noise improvement is relatively moderate. This is due to the fact that the coating mechanical loss gets worse thus cancelling some of the improvement coming from the temperature improvement.